# Superluminal propagation of evanescent modes as a quantum effect

Zhi-Yong Wang[1*], Cai-Dong Xiong[1], Bing He[2]

[1]School of Optoelectronic Information, University of Electronic Science and Technology of China, Chengdu 610054, CHINA
[2]Department of Physics and Astronomy, Hunter College of the City University of New York, 695 Park Avenue, New York, NY 10021, USA

*E-mail:   zywang@uestc.edu.cn

**Abstract**

Contrary to mechanical waves, the two-slit interference experiment of single photons shows that the behavior of classical electromagnetic waves corresponds to the quantum mechanical one of single photons, which is also different from the quantum-field-theory behavior such as the creations and annihilations of photons, the vacuum fluctuations, etc. Owing to a purely quantum effect, quantum tunneling particles including tunneling photons (evanescent modes) can propagate over a spacelike interval without destroying causality. With this picture we conclude that the superluminality of evanescent modes is a quantum mechanical rather than a classical phenomenon.

**Keywords**: Evanescent modes, superluminal propagation, quantum effect, virtual photons

**PACS**: 03.65.Xp, 41.20.Jb, 42.50.Nn, 03.65.Pm

## 1. Introduction

Nowadays, both theoretical and experimental investigations have presented a conclusion that the evanescent modes of the electromagnetic field can superluminally propagate [1-10]. At the level of quantum mechanics, via tunneling analogy the superluminal propagation of evanescent modes has been described as the quantum tunneling behavior of photons, which implies that the superluminality of evanescent modes is due to a quantum effect. In this paper, at the level of quantum field theory, we will further show that the superluminality of evanescent modes is due to a purely quantum effect, and clarify some misunderstandings on the physical properties of evanescent modes. As an application, we conclude that a recent objection [11-13] to the superluminality of evanescent modes is invalid, because the objection is completely based on classical mechanics by regarding electromagnetic waves as mechanical waves.

To avoid misunderstanding our argumentation (e.g., in spite of the genuine text of Refs. [9, 10], in Ref. [13] the author made improper claim that "they mistake a non-zero propagator for a non-zero commutator"), one should not confuse the following two issues: (1) whether a particle can propagate over a spacelike interval? (2) whether such propagation destroys causality (i.e., whether it means a measurement performed at one point can affect another measurement at a point separated from the first with a spacelike interval), if a particle does propagate over a spacelike interval? According to quantum field theory, a non-zero propagator or non-zero transition probability amplitude for a spacelike interval implies that a particle can propagate over the spacelike interval [14], but this spacelike propagation does not destroy causality provided that the commutator of two observables with a spacelike interval vanishes, that is, a measurement performed at one point does not affect another measurement at a point separated from the first with a spacelike interval.

On the other hand, the commutator between two field operators located at spacelike distance does not always vanish if the field operators are not observables. For example, in the Coulomb gauge, the commutator between electromagnetic potentials does not vanish for spacelike distances [15]. Moreover, in the quantization theory of evanescent modes [16], by assuming the high-frequency behavior of the refractive index, one can find that the commutator of evanescent field operators between two space-like separated points does not vanish, whose physical meaning and the related causality problem have been discussed by Stahlhofen and Nimtz [17].

## 2. Quantum field theory naturally explains superluminality of particles

A main reason to object all the existing theoretical and experimental investigations on the superluminality of evanescent modes lies in the fact that such superluminal propagation is in conflict with special relativity. However, special relativity has been developed on the basis of classical mechanics without taking into account any quantum-mechanical effect. On the other hand, because quantum field theory combines quantum mechanics with special relativity, such that it can give us such a conclusion [14, 18, 19]: owing to quantum-mechanical effect, a particle can propagate over a spacelike interval (without destroying Einstein's causality), which corresponds to the quantum tunneling phenomenon [18].

For example, just as S. Weinberg discussed [19] (with some different notations and conventions): "Although the relativity of temporal order raises no problems for classical physics, it plays a profound role in quantum theories. The uncertainty principle tells us that when we specify that a particle is at position $x_1$ at time $t_1$, we cannot also define its velocity precisely. In consequence there is a certain chance of a particle getting from $(t_1, \boldsymbol{x}_1)$ to $(t_2, \boldsymbol{x}_2)$ even if the spacetime interval is spacelike, that is, $|\boldsymbol{x}_1 - \boldsymbol{x}_2| > c|t_1 - t_2|$. To be more precise, the probability of a particle reaching $(t_2, \boldsymbol{x}_2)$ if it starts at $(t_1, \boldsymbol{x}_1)$ is nonnegligible as long as (we call Eq. (1) *Weinberg's formula*)

$$0 < (\boldsymbol{x}_1 - \boldsymbol{x}_2)^2 - c^2(t_1 - t_2)^2 \leq (\hbar/mc)^2, \tag{1}$$

where $\hbar$ is Planck's constant (divided by $2\pi$), c is the velocity of light in vacuum, and $m$ is the particle's mass, then $\hbar/mc$ is the particle's Compton wavelength. We are thus faced with our paradox: if one observer sees a particle emitted at $(t_1, \boldsymbol{x}_1)$, and absorbed at $(t_2, \boldsymbol{x}_2)$, and if $(\boldsymbol{x}_1 - \boldsymbol{x}_2)^2 - c^2(t_1 - t_2)^2$ is positive (but less than or equal to $(\hbar/mc)^2$), then a second observer may see the particle absorbed at $\boldsymbol{x}_2$ at a time $t_2$ before the time $t_1$ it is emitted at $\boldsymbol{x}_1$. There is only one known way out of this paradox. The second observer must see a particle emitted at $\boldsymbol{x}_2$ and absorbed at $\boldsymbol{x}_1$. But in general the particle seen by the second observer will then necessarily be different from that seen by the first observer (it is the antiparticle of the particle seen by the first observer)". In other words, to avoid a possible causality paradox, one can resort to the particle-antiparticle symmetry. The process of a particle created at $(t_1, \boldsymbol{x}_1)$ and annihilated at $(t_2, \boldsymbol{x}_2)$ as observed in a frame of reference, is identical with that of an antiparticle created at $(t_2, \boldsymbol{x}_2)$ and annihilated at $(t_1, \boldsymbol{x}_1)$ as observed in another frame of reference.

In fact, Weinberg's argument is equivalent to the usual argument in quantum-field-theory textbooks: let $\phi(x)$ stand for a scalar field operator, $|0\rangle$ denote the vacuum state, then $\langle 0|\phi(x)\phi(y)|0\rangle$ represents the transition probability amplitude from the state $\phi(y)|0\rangle$ to the state $\phi(x)|0\rangle$ [14], such that $|\langle 0|\phi(x)\phi(y)|0\rangle|^2$ corresponds to the probability for a scalar particle to propagate over the spacetime interval $(x-y)^2$. In particular, if the probability amplitude for a scalar particle propagating over a spacelike interval $(x-y)^2 < 0$ is denoted as $D(x-y) = \langle 0|\phi(x)\phi(y)|0\rangle$, then according to quantum field theory, $D(y-x) = \langle 0|\phi(y)\phi(x)|0\rangle$ represents the probability amplitude for the corresponding antiparticle propagating backwards over the spacelike interval. Because $D(x-y) = D(y-x)$ for $(x-y)^2 < 0$, the two spacelike processes are undistinguishable and the commutator $[\phi(x),\phi(y)] = \langle 0|[\phi(x),\phi(y)]|0\rangle = 0$, such that the causality is maintained. Therefore, Weinberg has provided another way of looking at the statement that "a measurement performed at one point does not affect another measurement at a point separated from the first with a spacelike interval". Studying a quantum Lorentz transformation one can also obtain Eq. (1) [20].

In fact, Eq. (1) is just an approximation of a more rigorous result. For our purpose, let us derive the rigorous result within the framework of quantum field theory. The transition probability amplitude for any particle to propagate over the spacetime interval $(x-y)^2$ can be expressed in terms of $D(x-y) = \langle 0|\phi(x)\phi(y)|0\rangle$, and then we can ignore the spin degree of freedom and take the scalar field $\phi(x)$ for example. For convenience let $y = (0,0,0,0)$, $x = (t,r,0,0)$, and denote $D(t,r) = D(x-y)$, according to quantum field theory, one has (up to a constant factor)

$$D(t,r) = \int_{-\infty}^{+\infty} \frac{dp}{2\pi} \frac{c}{2E_p} \exp[-i(E_p t - pr)/\hbar], \qquad (2)$$

where $E_p = \sqrt{p^2c^2 + m^2c^4}$. Let $H_0^{(2)}(z)$ denote the zero-order Hankel function of the second kind, as the spacetime interval is spacelike (i.e., $c^2t^2 - r^2 < 0$), one can prove that,

$$D(t,r) = (-i/4)H_0^{(2)}(-i\sqrt{r^2 - c^2t^2}/\lambdabar), \qquad (3)$$

where $\lambdabar = \hbar/mc$ is the Compton wavelength. Therefore, the asymptotic behaviors of $D(t,r)$ are governed by the Hankel function of imaginary argument: $D(t,r)$ falls off like $\sqrt{1/z}\exp(-z)$ for $z = \sqrt{r^2 - c^2t^2}/\lambdabar \to +\infty$, while falls off faster than $\sqrt{1/z}\exp(-z)$ for the other $z = \sqrt{r^2 - c^2t^2}/\lambdabar$. In the observable sense, $D(t,r)$ is always ignored for $z = \sqrt{r^2 - c^2t^2}/\lambdabar > 1$, that is, one always takes the approximate as follows:

$$D(t,r)\begin{cases} = 0, \text{ for } c^2t^2 - r^2 < -(\hbar/mc)^2 = -\lambdabar^2 \\ \neq 0, \text{ for } 0 > c^2t^2 - r^2 \geq -(\hbar/mc)^2 = -\lambdabar^2 \end{cases}. \quad (4)$$

According to the approximate given by Eq. (4), for the spacelike interval $c^2t^2 - r^2 < 0$, the probability amplitude $D(t,r)$ for the particle to propagate from $y = (0,0,0,0)$ to $x = (t,r,0,0)$ is nonnegligible as long as the Weinberg's formula given by Eq. (1) is satisfied. In other words, the Weinberg's formula given by Eq. (1) is just an approximate of the rigorous result given by Eq. (3). The rigorous result (3) implies that there are in principle no limitations to the spacelike interval of $\sqrt{r^2 - c^2t^2}$, but the probability $|D(t,r)|^2$ falls off rapidly for large $\sqrt{r^2 - c^2t^2}$, and the spacelike process cannot be observed provided that the probability is too small.

Taking an electron for example, the Compton wavelength of the electron is $\hbar/mc \approx 3.87 \times 10^{-10}$ (millimeter, mm). For any spacelike interval there is always an inertial reference in which one has $t_1 = t_2$, using Eq. (1) one has

$$\sqrt{(\boldsymbol{x}_1 - \boldsymbol{x}_2)^2 - c^2(t_1 - t_2)^2} = |\boldsymbol{x}_1 - \boldsymbol{x}_2| \leq \hbar/mc \approx 3.87 \times 10^{-10} \text{ mm}. \quad (5)$$

The spacelike process occurring within such a spatial region is difficult to be observed. In fact, within the Compton wavelength of the electron, the many-particle effects arising from the creations and annihilations of virtual electron-positron pairs cannot be ignored. On the other hand, for large spacelike interval the corresponding probability becomes so small that the spacelike process cannot be observed. In a word, Eq. (1) as the approximate of the rigorous result (3), describes a spacelike process with a sufficiently large probability (in the observable sense).

**3. Superluminal propagation of evanescent modes as a quantum effect**

Photons inside a hollow waveguide can be treated as free massive particles with an effective mass $m_{\text{eff}} = \hbar\omega_c/c^2$ [9, 21], where $\omega_c$ is the cut-off frequency of the waveguide, then the argument in Section 2 is also valid for the guided photons. In fact, quantum field theory tells us that [9, 10]: owing to a quantum effect, photons inside a waveguide can propagate over a spacelike interval, which corresponds to the fact that the evanescent modes can propagate superluminally through an undersized waveguide (i.e., the photonic tunneling phenomenon). Likewise, here the Einstein causality is preserved via the particle-antiparticle symmetry, but for the moment the antiparticle of a photon is the photon itself, such that the process that a photon propagates superluminally from A to B as observed in an inertial frame of reference, is equivalent to that the photon propagates superluminally from B to A as observed in another inertial frame of reference. Moreover, via quantum Lorentz transformation [20] or by developing special relativity on the basis of quantum mechanics [21], another theoretical evidence for the superluminality of evanescent modes can be obtained, which also shows that the superluminal behavior of evanescent modes arises from a quantum effect, i.e., the Heisenberg's uncertainty. However, here the theoretical evidence is obtained at the level of quantum mechanics, it is just an approximate of the more rigorous result given by Eq. (3), i.e., the related spacelike interval in a spacelike process is just the one with a sufficiently large propagation probability (i.e., with $|D(t,r)|^2 \geq (1/16)[H_0^{(2)}(-i)]^2$ ).

As an example, let the cut-off frequency of a waveguide be $\omega_c = 9.49 \text{GHz}$, the effective Compton wavelength of photons inside the waveguide is $\hbar/m_{\text{eff}}c = c/\omega_c \approx 31.6\text{mm}$, which is far too larger than $3.87 \times 10^{-10}$ mm (i.e., the Compton wavelength of the electron). That is, for tunneling photons, the spacelike interval with a sufficiently large propagation probability is

$$\sqrt{(\boldsymbol{x}_1 - \boldsymbol{x}_2)^2 - c^2(t_1 - t_2)^2} \leq \hbar/m_{\text{eff}}c = c/\omega_c \approx 31.6\text{mm}. \qquad (6)$$

Therefore, contrary to the superluminal behavior of tunneling electrons, the superluminal behavior of evanescent modes can be easily observed experimentally. It is important to mention that, Eq. (6) does not conflict with those experimental results with the largest tunneling distance larger than 31.6 mm, because: 1) for a given spacelike interval $\sqrt{r^2 - c^2 t^2}$, the propagation distance $r = |\boldsymbol{x}_1 - \boldsymbol{x}_2|$ is related to the propagation time $t = (t_1 - t_2)$; 2) Eq. (6) as an approximate of the more rigorous result given by Eq. (3), just corresponds to the result with a sufficiently large propagation probability (i.e., satisfying $|D(t,r)|^2 \geq (1/16)[H_0^{(2)}(-i)]^2$ ), while Eq. (3) tells us that there are in principle no limitations to the spacelike interval, though the probability $|D(t,r)|^2$ falls off rapidly for large spacelike interval.

Moreover, contrary to electrons, photons are bosons and do not carry any charge, by increasing the number of tunneling photons the observable spacelike interval can be augmented *ad lib*. Eq. (3) shows that the probability amplitude $D(t,r)$ falls off exponentially (but does not vanish) as the spacelike interval $\sqrt{r^2 - c^2 t^2} \to +\infty$; on the other hand, for the spacelike interval there is always an inertial reference in which one has $t=0$. Therefore, even if the propagation distance $r \to +\infty$, the propagation time can be arbitrarily small, which is in agreement with the Hartman effect [22].

In frustrated total internal reflection, evanescent modes as near field consist of virtual photons [23-24], these virtual photons correspond to the elementary excitations of electromagnetic interactions. Now we show that evanescent modes inside an undersized waveguide are also identical with virtual photons. As we know, the near fields of a dipole antenna fall off with the distance $r$ from the antenna like $1/r^n$ ($n \geq 2$). However, if we assume that an aerial array formed by an infinite set of infinite-length line sources arranging in a periodic manner (with the period $r_0$), then the near fields of the aerial array falls off like $\exp(-r/r_0)$. With respect to the TE$_{10}$ mode, an undersized waveguide is equivalent to such an aerial array [25] and evanescent modes inside the undersized waveguide are equivalent to the near fields of the aerial array, which implies that evanescent fields inside the waveguide can also be described by virtual photons. As we know, the propagation of virtual photons is due to a purely quantum-mechanical effect, which also implies that one cannot understand the propagation of evanescent waves via classical mechanics. To show the evanescent TE$_{10}$ mode (with the frequency $\omega < \omega_c$, where $\omega_c$ is the cut-off frequency) is equivalent to the near field of the aerial array, basing on Ref. [25], one ought to make the following replacements:

$$\omega_c = \sqrt{\omega_c^2 - 0^2} \to \sqrt{\omega_c^2 - \omega^2}, \quad \exp(i \cdot 0 \cdot t) = 1 \to \exp(i\omega t), \qquad (7)$$

$$(\frac{\partial^2}{\partial x^2} + \frac{\partial^2}{\partial z^2})E_y(x,z) = 0 \to (\frac{\partial^2}{\partial x^2} + \frac{\partial^2}{\partial z^2} - \frac{\partial^2}{c^2 \partial t^2})E_y(x,z,t) = 0, \qquad (8)$$

$$E_y(x,z) = E_0 \sin(\frac{\pi x}{a}) \exp(-\frac{\omega_c z}{c}) \rightarrow E_y(x,z,t) = E_0 \sin(\frac{\pi x}{a}) \exp(i\omega t - \kappa z), \quad (9)$$

where $z_0 = a/\pi = c/\omega_c$, and $\kappa = \sqrt{\omega_c^2 - \omega^2}/c$. The presence of the decay factor $\exp(-\kappa z)$ implies that the field $E_y(x,z,t)$ mainly exists in the neighborhood of the aerial array, and then is the near field of the aerial array.

**4. Some misunderstandings about the physical properties of evanescent modes**

To show that the superluminal propagation of evanescent modes is due to a purely quantum effect, in addition to the argument presented above, some misunderstandings on the physics properties of evanescent modes should be clarified. In particular, these misunderstandings appear in the objection [11-13] to the superluminality of evanescent modes.

Firstly, classical electromagnetic waves are conceptually different from mechanical waves such as water waves, acoustic waves, waves on a string, etc. (e.g., only via media can mechanical waves propagate, while electromagnetic waves can propagate in vacuum). For example, the wave nature of mechanical waves is usually described by classical mechanics, while the wave-particle dualism of light tells us that the wave nature of classical electromagnetic waves is essentially a quantum mechanical issue, and historically quantum mechanics arises from extending the wave-particle dualism of light to that of massive particles. In terms of the spinor representation of electromagnetic field [26, 27], one can obtain the quantum-mechanical theory of single photons (note that the usual quantum theory of electromagnetic field is referred to the field-quantized theory rather than quantum mechanics). In fact, to interpret the two-slit interference experiment of single photons one-by-one emitted from a light source, one has to regard the behaviors of classical electromagnetic waves as the quantum-mechanical ones of single photons, and here have nothing to do with the quantum-field-theory effects such as the creations and annihilations of photons, or the zero-point fluctuations of quantum electromagnetic field.

From the point of view of classical mechanics, inside an undersized waveguide evanescent waves have support everywhere (through exponential damping) along the undersized waveguide, and "the propagation of evanescent modes" is not a well-defined concept. However, this classical picture just provides us with a phenomenological description. From the point of view of quantum mechanics, now that some fraction of an electromagnetic wave beam entering in the input side of an undersized waveguide with finite length will come out of the exit of the waveguide, it indicates that there must have a physical process that some photons in the evanescent wave beam propagate through the undersized waveguide. As a quantum tunneling phenomenon, this physical process is due to a purely quantum-mechanical effect without any classical correspondence, such that only via quantum theory can one explain the propagation of evanescent modes. In other words, if evanescent waves could not propagate, likewise any other quantum tunneling phenomenon could not occur, because the wavefunctions of tunneling particles within a potential barrier are similar to evanescent electromagnetic waves: they possess imaginary wave-numbers; they do not describe propagating waves but evanescent waves.

On the other hand, plug in a signal into a tunnel and as long as it can be read out at the other end, Einstein causality is violated [17].

In the objection to the superluminality of evanescent modes, evanescent waves are by mistake regarded as "exponentially attenuated standing waves". To clarify such a misunderstanding, let us assume that a hollow waveguide is placed along the direction of z-axes, and the waveguide is a straight rectangular pipe with the transversal dimensions $a$ and $b$ ($a > b$, the cross-section of the waveguide lies in $0 \leq x \leq a$ and $0 \leq y \leq b$). Inside the waveguide, for electromagnetic waves with the frequency $\omega$ and wave-number vector $\boldsymbol{k} = (k_x, k_y, k_z)$, take the electric field component $E_x$ for example, it can be written as

$$E_x(x,y,z,t) = A\cos(k_x x)\sin(k_y y)\exp(i\omega t - ik_z z), \qquad (10)$$

where $A$ is a constant factor, $k_x = n\pi/a$ and $k_y = l\pi/b$ ($n, l = 0, 1, 2, ...$) are the wavenumbers along the x-axis and y-axis directions, respectively. In Eq. (10), $f(x,y) \equiv A\cos(k_x x)\sin(k_y y)$ represents a standing-wave factor (in which the wavenumbers $k_x$ and $k_y$ are real), which implies that the electromagnetic waves inside the waveguide form a standing-wave structure along the $xy$ plane (i.e., along the cross-section of the waveguide) for both propagation and evanescent modes. On the other hand, in Eq. (10), for a real $k_z$, $\exp(i\omega t - ik_z z)$ is a propagation factor and then $E_x(x,y,z,t)$ represents propagation modes; while for an imaginary $k_z = -i\kappa$ ($\kappa$ is a real number), $\exp(i\omega t - ik_z z)$ becomes the attenuation factor of $\exp(i\omega t - \kappa z)$ and then $E_x(x,y,z,t)$ represents evanescent modes. Therefore, along the z-axis direction (i.e., along direction of the waveguide), the electromagnetic waves form propagating- and evanescent-wave structures for the propagation and evanescent modes, respectively. In other words, along the direction of the waveguide, there is no standing-wave structure, such that the evanescent modes cannot be regarded as "exponentially attenuated standing waves". If one insists on regarding the evanescent modes as "exponentially attenuated standing waves" for the reason that they contain the standing-wave factor $f(x,y) \equiv A\cos(k_x x)\sin(k_y y)$, he would have to call the propagation modes "propagating standing waves", which is a self-contradictory appellation.

A reason for denying that evanescent modes have quantum mechanical behaviors is that "evanescent modes can completely be described by Maxwell's equations". However, it is well known that Maxwell's equations not only describe the classical electromagnetic field, but also the quantum one (at the level of quantum field theory). On the other hand, just as mentioned above, the two-slit interference experiment of single photons shows that the behavior of classical electromagnetic waves corresponds to the quantum mechanical one of single photons (in the first-quantized sense).

## 5. Conclusions

The two-slit interference experiment of single photons tells us that, contrary to mechanical waves described by classical mechanics, the wave nature of electromagnetic waves is essentially a quantum mechanical issue. Just as all other quantum tunneling phenomena, the propagation of evanescent modes attributes to the quantum-mechanical behavior of photons and cannot be understood via classical mechanics. The superluminal propagation of evanescent modes can be interpreted by the quantum-mechanical behavior of single photons (in terms of photonic quantum tunneling), or by the quantum-field-theory behavior of the electromagnetic field (in terms of a non-zero transition probability amplitude for a spacelike interval, or in terms of spacelike virtual photons). In a word, the superluminal propagation of evanescent modes is a purely quantum mechanical phenomenon without any classical correspondence. As a result, any objection to the superluminality of evanescent modes is invalid provided that the objection is completely based on classical mechanics.

**Acknowledgments**

The first author (Z. Y. Wang) would like to thank Professor G. Nimtz for his many helpful discussions. This work was supported by the National Natural Science Foundation of China (Grant No. 60671030) and Project supported by the Scientific Research Starting Foundation for Outstanding Graduate, UESTC , China (Grant No. Y02002010501022).